\documentclass[a4paper, amsfonts, amssymb, amsmath, reprint, showpacs, showkeys, nofootinbib, twoside, floatfix]{revtex4-1}
\usepackage[english]{babel}
\usepackage[utf8]{inputenc}
\usepackage{upgreek}
\usepackage[colorinlistoftodos, color=green!40, prependcaption]{todonotes}
\usepackage{amsthm}
\usepackage{mathtools}
\usepackage{physics}
\usepackage{siunitx}
\usepackage{xcolor}
\usepackage{graphicx}
\usepackage[autostyle]{csquotes}
\usepackage[left=16mm,right=16mm,top=35mm,bottom=20mm,columnsep=14pt]{geometry} 
\usepackage{adjustbox}
\usepackage{placeins}
\usepackage[T1]{fontenc}
\usepackage{lipsum}
\usepackage{url}
\usepackage{natbib,hyperref}
\hypersetup{colorlinks=true, urlcolor=blue, linkcolor=blue, citecolor=blue}
\usepackage{multirow, booktabs}
\usepackage{epstopdf}

\begin{document}
\renewcommand{\figurename}{Fig.{}}
\title{Effect of slicing in velocity map imaging for the study of dissociation dynamics}

\author{Narayan Kundu\textit{$^{1}$}}
\author{Dipayan Biswas\textit{$^{1}$}}
\author{Vikrant Kumar\textit{$^{1}$}}
\author{Anirban Paul\textit{$^{1}$}}

\author{Dhananjay Nandi\textit{$^{1,2}$}}%
 \email{dhananjay@iiserkol.ac.in}
\affiliation{$^1$Indian Institute of Science Education and Research Kolkata, Mohanpur-741246, India.\\ $^2$Center for Atomic, Molecular and Optical Sciences $\&$ Technologies, Joint initiative of IIT Tirupati $\&$ IISER Tirupati, Yerpedu, 517619, Andhra Pradesh, India}

\begin{abstract}
Inelastic collision dynamics between isolated gas-phase carbon monoxide molecules and low energetic electrons (< 50 eV) has been studied using state-of-the-art velocity map imaging apparatus and reported previously \cite{chakraborty2016dipolar,nag2015fragmentation}. These were based on data analysis using the time-gated parallel slicing technique, which has recently revealed the drawback of lower momentum ion exaggeration mainly due to the inclusion of whole Newton sphere's of diameter $\le$ parallel slicing time window. To overcome this drawback, we report implementing a wedge slicing technique so that every momentum sphere contributes equally to the statistics. We also present a comparative study between these two techniques by reanalyzing the data using the time-gated parallel slicing technique. Unlike parallel slicing, the wedge slicing technique better represents the dissociation dynamics, particularly for the ions with low kinetic energy.

\end{abstract}

\maketitle

\section{Introduction} 
A well-resolved ion image is nothing but good optimization of its positional parameters. Position-sensitive detection of the nascent products has witnessed a tremendous amount of popularity in the last decade \cite{whitaker2003imaging}. Applying suitable techniques to the extracted data set can help uncover the physics behind the dynamics. In the realm of atomic and molecular physics, the velocity map imaging (VMI) technique has several advantages over the conventional turntable technique in determining the speed and angular distribution of fragmented products \cite{eppink1997velocity,thoman1988two}. VMI is revealing its fruitfulness to the ion imaging \cite{krishnakumar2018symmetry,lee2020three,neumark2008slow,sarma2012inelastic,suits2001imaging,ghafur2009velocity,wester2014velocity} for molecular physics, physical chemistry, chemical physics, plasma physics, radiation therapy, electron-induced chemistry, biochemistry, optics and photonics etc.

Ion imaging is an indispensable experimental technique, mainly used to measure the velocity and angular distribution of the fragment ions or particles produced by different molecular reactions ($e.g.$ photodissociation, photoionization, dissociative electron attachment). In 1987, Chandler and Houston \cite{chandler1987two} implemented this technique on the photodissociation dynamics of methyl iodide (CH$_3$I). Later, in 1997, Eppink and Parker \cite{eppink1997velocity} introduced the velocity map imaging technique in which they employed electrostatic lenses to focus all the ions having the same velocity vector at a single point on the detector, which improved the kinetic energy resolution of the fragments. The capability of probing the full three-dimensional velocity and angular distribution of scattered particles has made this technique unique. In the earlier days, the three-dimensional Newton sphere of the nascent fragments was constructed from the two-dimensional data using the pBASEX Abel inversion algorithm by direct sampling of the basis functions on a Cartesian grid or by rebinning the data into polar coordinates that generate artificial noise along the axis of symmetry \cite{garcia2004two}. Using delayed pulsed extraction, Gebhardt \textit{et al.} \cite{gebhardt2001slice} detects the Newton sphere with sufficient velocity spread. This spread is in good agreement with the spatial resolution of the fragment ions. After that, a narrow detector time gate (< 40 ns) is applied to record the central slice of the Newton sphere. Later, Nandi and Krishnakumar \cite{nandi2005velocity} implement this technique to measure the kinetic energy and angular distribution of the anionic fragments. They recorded the time of arrival of each ion along with the position of hits on the 2D-PSD. The more informative central slice could be obtained from slices obtained using parallel time-gate to the TOF values during offline analysis. However, the outcome from this time-gated two-dimensional velocity slice image (VSI) is equivalent to that obtained from inverse Abel transformations (IAT) which take a two-dimensional projection and reconstruct a slice of the cylindrically symmetric three-dimensional distribution. Later, in 2005, Li \textit{et. al.} \cite{li2005megapixel} report megapixel ion imaging implementing standard video and dc slicing technique. Recently, Dick bypasses the Abel inversion technique by introducing the maximum entropy concept to reconstruct the velocity maps from its ion images. In 2014, Dick introduced \cite{dick2014inverting} maximum entropy velocity image reconstruction (MEVIR), directly analogous to Abel inversion and maximum entropy velocity Legendre reconstruction (MEVELER). MEVELER provides notably better than MEVIR for ion images with low intensity. Later, in 2019, Dick implemented maximum entropy Legendre expanded image reconstruction (MELEXIR) \cite{dick2019melexir}, a similar technique like MEVELER but with several advantages. It uses low CPU memory, is faster, and can analyze the data containing higher-order and odd-order Legendre polynomials. To study weak vibronic coupling, Neumark \cite{neumark2008slow} developed the technique of slow electron VMI (SEVI). 

In this article, we have implemented a wedge slicing technique for analyzing the dissociative electron attachment (DEA) and ion-pair dissociation (IPD) dynamics to overcome the difficulties that arise during the same dynamics study using the time-gated parallel slicing method. In this comparative study, collision with carbon monoxide is the center of attention due to its attractive features in low-energy electron collision.

\section{Instrumentations and Slicings}
The experiment is based on the cross-collision between electron and molecular beam, which is carried out in a vacuum chamber, pumped down up to $5\times10^{-10}$ mbar using a turbomolecular vacuum pump. The measurements are done by the highly differential state of art VMI spectrometer specially built to study low-energy electron-molecule interactions. These have already been discussed in detail elsewhere \cite{nag2015complete}. The pulsed electron beam interacts with an effusive molecular beam formed by a capillary tube. After a finite time delay, the fragmented ions are extracted using an extraction pulse into a TOF-based three-field mass spectrometer. When the low energetic electron interacts with the molecule, each event yields two partner fragments flying with equal momentum in opposite directions in the center of the mass frame. When we repeat the same event many times, spherical distributions of fragments will be built up in the velocity space, commonly known as the Newton sphere. The size of this Newton Sphere tells us about the balance of internal and transnational energy in the reaction \cite{whitaker2003imaging}. Newton sphere contains kinematically complete pieces of information for a given scattering process. Fragments with the same velocity vector are mapped to a single point on the 2D PSD, consisting of three microchannel plates (MCP) in a Z-stack configuration. The encoding of the positions of hits (x and y) is done by a three-layer delay line hexanode \cite{jagutzki2002multiple}, placed behind the MCPs. The TOF of the detected ions is then calculated using the back-MCP signals. We use the CoboldPC software \cite{ullmann1999list} to save the x \& y coordinates along with the flight time (t) of all detected ions in a list-mode format (LMF). Others experimental details are briefly discussed in our previous reports \cite{nag2015fragmentation,chakraborty2016dipolar}. 
\begin{figure}[hbt]
  \includegraphics[scale=0.48]{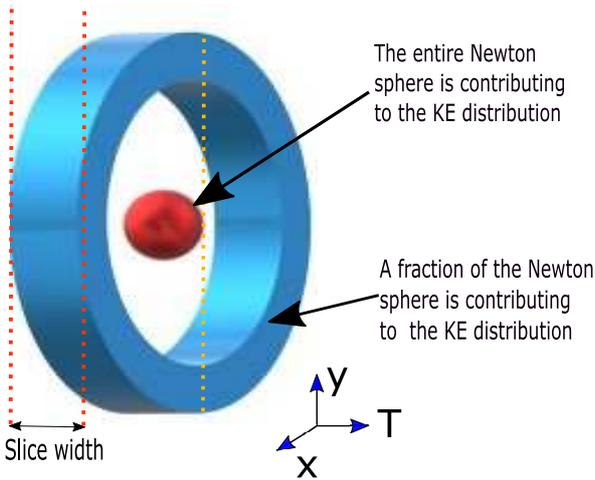}
  \caption{\footnotesize{A schematic diagram of time-gated parallel slice is shown here. The ions with lower momentum in red color contributes more to the kinetic energy distribution than the ion with higher momentum as shown in sky blue color.}}
  \label{fig:flat_slice}
\end{figure}
\begin{figure}[hbt]
  \includegraphics[scale=0.13]{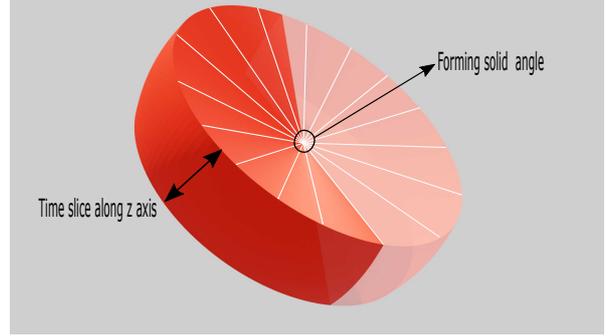}
  \caption{\footnotesize{A schematic three-dimensional representation of wedge slice. The TOF is scaled linearly to construct the z-axis, perpendicular to the detector plane. Half part of this wedge shape slice has given deep red color and other half as light red to achieve better visualisation effect.}}
  \label{fig:wedge_slice}
\end{figure}

Previously, the raw data was recorded in list mode format (LMF) only and extracted time-gated parallel slices were used to analyze the entire dynamics. The TOF of each ion is measured as the time interval between the delayed start pulse and the detected ion signal. Let the spectrometer axis is behaving as a z-axis perpendicular to the 2D-PSD plane (x-y plane). So, the TOF of the detected anions is an indirect measurement for the z-component of the momentum. Now, the raw data is available in a text file that can be sliced externally as desired. In a TOF mass spectrometer, the TOF of the fragmented ions is roughly proportional to the square root of the charge to mass ratio \cite{wiley1955time,mamyrin2001time}. In our experiment, we are considering the nascent fragments with one unit of negative charge. Hence, using the proper mass calibration (calibrated with O$^{-}$/O${_2}$), we have resolved the mass of the fragment anions (O$^{-}$/CO). At 10.5 eV incident electron energy, typical FWHM for the TOF spectra of O$^{-}$/CO is 400 ns. 

\begin{figure*}
 \centering
  \includegraphics[scale=0.9]{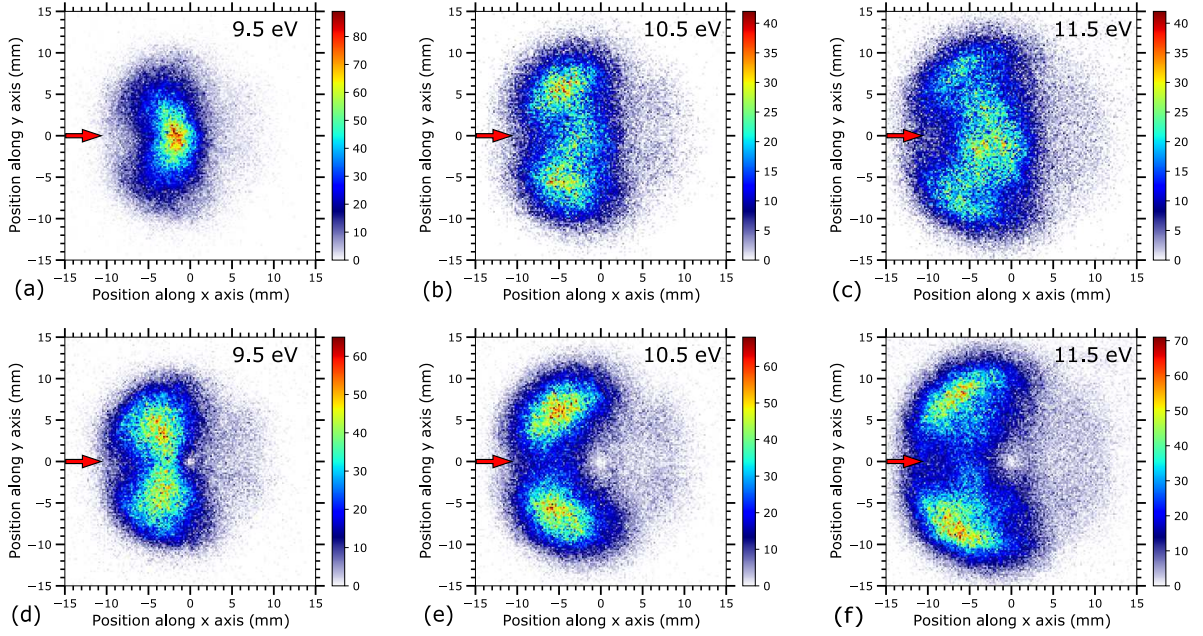}
  \caption{\footnotesize{a,b and c are the 50 ns parallel time-sliced images of O$^{-}/$CO at 9.5, 10.5 and 11.5 eV incident electron energies, respectively.d,e and f are the wedge slice images of O$^{-}/$CO at 9.5, 10.5 and 11.5 eV incident electron energies, respectively. The red arrow indicates the electron beam direction. Wedge slices are considered in such a way that the maximum time interval along the z-axis approximately lies 50 ns.}}
  \label{fig:covsi_dea}
\end{figure*} 
In the most usual and conventional slicing technique, a reasonably narrow time window is selected around the central time of flight of each ion as shown in Fig. \ref{fig:flat_slice}. The central TOF corresponds to the equatorial plane of the ellipsoid formed in the (x,y,t) coordinate system. Using a suitable scale factor, we can convert the TOF value to the z-component of momentum that eventually constructs the 3D momentum sphere where the central TOF should correspond to z = $0$ plane. A narrow TOF window around the central TOF allows us to construct a narrow parallel time slice at averaging z = $0$ plane enabling us to eliminate the $z^2$ term in the momentum distribution (p=$\sqrt{p_{x}^{2}+p_{y}^2}$). The kinetic energy of the ions can be calculated as E $=$ cr$^2$, where c is a mass-independent calibration constant, and r is the radius vector of the more informative central slice (biggest in radius). In VMI, fragments with specific energy map onto the detector like concentric circles with a fixed radius, irrespective of the masses of the detected fragments. The beauty of such a state of art mapping is that mass differences among different nascent fragments are balanced by the time of flight of those fragments.
\begin{figure*}%
    \centering
    \includegraphics[scale=0.85]{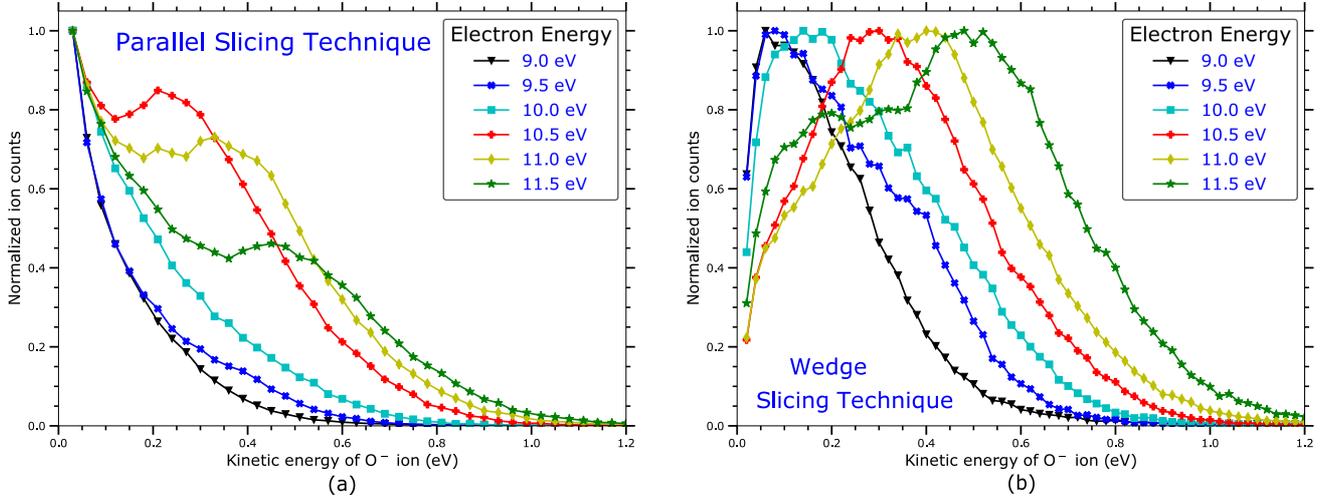}
  \caption{\footnotesize{ (a) Kinetic energy distributions of O$^-$ ions are extracted using time-gated (50 ns) parallel slicing technique as a function of incident electron energy. (b) kinetic energy distributions of O$^-$ ions are extracted using wedge slicing technique for different incident electron energy as shown in legend. }}
  \label{fig:co_ke_dea}
\end{figure*}
The time-gated parallel slicing technique works unconditionally fine if no low energetic nascent fragments are present in its Newton sphere. In the case of fragments that have a reasonable contribution from low energetic ions, the parallel slicing technique seriously exaggerates the contribution of low energetic ions than its high energetic ions \cite{moradmand_wedge}. Recently Nag \textit{et al.} \cite{nag_nccn} has obtained kinetic energy of the fragments using Newton half-sphere. It provides an excellent result in kinetic energy determination provided that the detected Newton half-sphere is well enough to maintain the VMI condition. To overcome this drawback of time-gated parallel slicing, a constant elevation in 2$\pi$ azimuth angle is considered from the sphere's center, conventionalized as a wedge slice or weighted solid angle slice. A schematic representation of a wedge slice is shown in Fig. \ref{fig:wedge_slice}. Here, the elevation angle ($\lambda$) is chosen so that the maximum time slice is about 50 ns, a comparable width to the parallel time gate. Firstly, we have used a linear scale factor on TOF to construct the Newton sphere with a radius numerically equal to that of the parallel central slice. Instead of implementing the Newton half-sphere technique, we go for the wedge slicing method as we notice a slight deflection from the VMI condition for the entire Newton sphere. Long TOF (> 4000 ns) and well spread Newton sphere (FWHM $\neq$ 400 ns) could be a rationale to observe a slight distortion of VMI in y-t momentum slice images. Thus, we have considered a statistically better wedge slice containing a part of the time-gated central slice.       
\begin{figure}
  \includegraphics[scale=0.35]{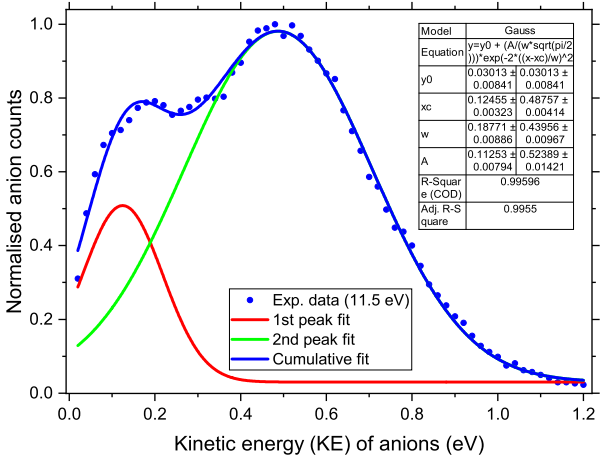}
  \caption{\footnotesize{The kinetic energy distribution of O$^-$ ions at 11.5 eV incident electron energy. The distribution is fitted using two peaks Gauss model and the obtained fit parameters are given within the figure.}}
  \label{fig:ke_fit}
\end{figure}
\begin{figure*}
    \centering
      \includegraphics[scale=0.85]{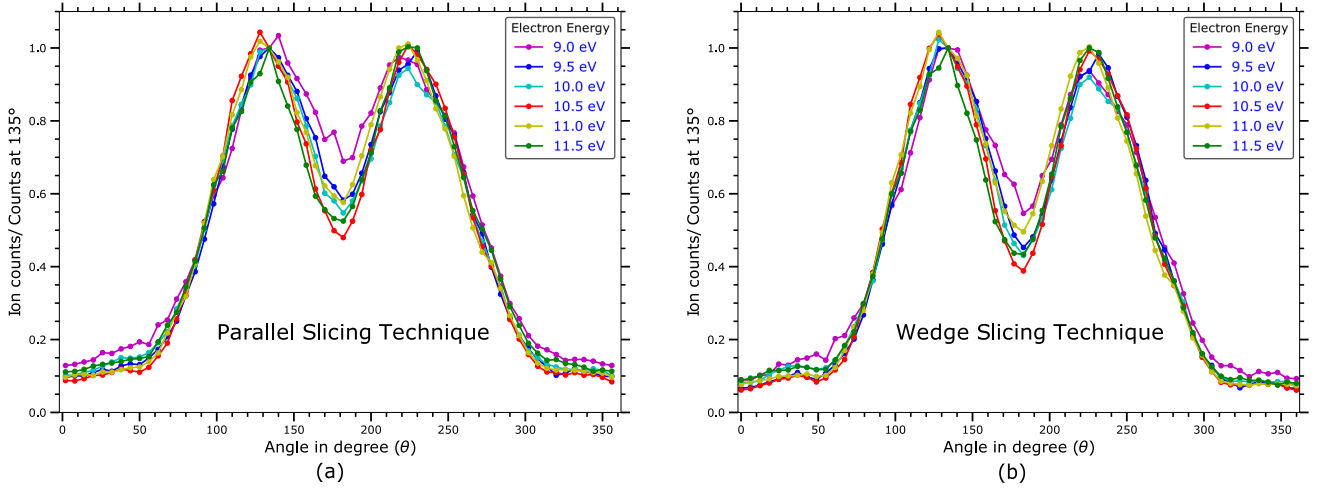}
  \caption{\footnotesize{(a) Angular distributions of fragmented O$^{-}$ ions have been extracted using time-gated (50 ns) parallel slicing technique as a function of incident electron energy. (b) Angular distributions of fragmented O$^{-}$ ions have been extracted using the wedge technique as a function of incident electron energy. All the angular distributions are normalised at angle 135$^{\circ}$ with respect to the electron beam direction.}}
  \label{fig:co_ad_flat_wedge_all}
\end{figure*}
Extracting kinetic energy distribution, the wedge slice or Newton half-sphere analysis is susceptible to the scaling factor used for TOF to z-momentum conversion. In wedge slicing, the kinetic energy of the fragments is calculated as E=cr$^2$=c(x$^2$+y$^2$) and the energy calibration constant C is evaluated using O$^{-}$/O${_2}$.
\section{Results and Discussions} 
We have reanalyzed the dynamics of CO for dissociative electron attachment (DEA) and ion-pair dissociation (IPD) processes using both the parallel and wedge slicing techniques. DEA, a two-step resonance process where an incident low energy electron (< 15 eV) attaches to the molecule and forms a temporary negative ion (TNI) state. These TNIs subsequently dissociate to a negative ion fragment and one or more than one neutral fragment. But, in the intermediate range of electron energy (15-50 eV), another process called ion-pair dissociation (IPD) or dipolar dissociation (DD) can initiate. Unlike DEA, the electron does not get captured by the target molecule in IPD/DD. Instead, the electron transfers a part of its kinetic energy to the molecule resulting in the excitation of the molecule beyond its first ionization potential and subsequently dissociates into cation and anion fragments \cite{suits2006ion}.
\subsection{DEA to carbon monoxide}
The DEA dynamics of carbon monoxide elegantly reveal the nature of forward-backward asymmetry as reported by Nag \textit{et al.} \cite{nag2015fragmentation}. Fig. \ref{fig:covsi_dea} shows velocity slice images (VSI) of O$^-/$CO for both the parallel and wedge slicing techniques at 9.5, 10.5 and 11.5 eV incident electron energies, respectively. Each VSI image has a 50 ns time window for the parallel slicing technique. In the case of the wedge slicing method, 10$^{\circ}$ elevation angle is set up corresponding to a maximum 50 ns time-slicing. If we carefully look at the VSI images for all the incident electron energies, especially at 9.5 eV, then the exaggerated nature near the center of the sliced image is noticeable. This statement is a rationale for finding the same angular distribution (low intensity along the beam axis) analyzed using both the techniques, as shown in Fig. \ref{fig:covsi_dea}. However, the weak contribution to the forward lobe is visible in the wedge slicing technique also. Inspecting the parallel slice image at 9.5 eV, one can assume that the maximum intensity lies near about 0$^{\circ}$ or 180$^{\circ}$ beam direction, which is resolved in the wedge slicing method.
\begin{figure*}%
  \centering
  \includegraphics[scale=0.85]{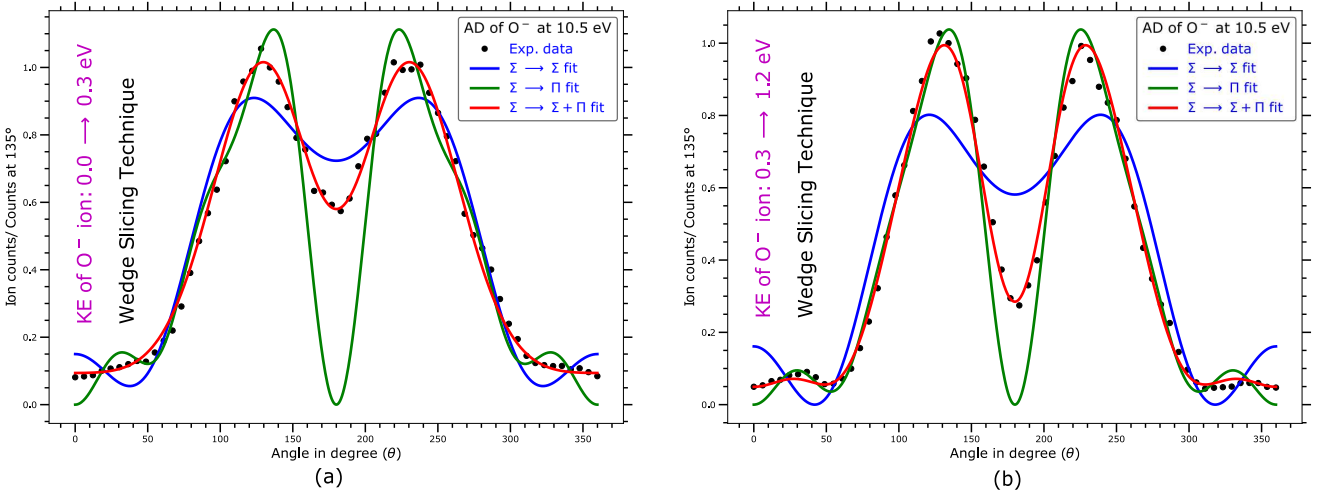}
  \caption{\footnotesize{ The angular distribution of O$^{-}/$CO at 10.5 eV incident electron energy is extracted using (a) time-gated (50 ns) parallel slicing method and (b) Wedge technique. Both distributions have been fitted using the model given by O'Malley and Taylor \cite{o1968angular} for initial state associated with $\Sigma$ symmetry to final states associated with $\Sigma$, $\Pi$ and $\Sigma$ $+$ $\Pi$ symmetries.}}
  \label{fig:ad_co_10_5_fit}
\end{figure*}
\begin{table*}
 \centering
\caption{The model parameters reported here for the angular distribution of O$^-$ ions with kinetic energy 0.0 to 0.3 eV at 10.5 eV incident electron energy. The a$_m$’s and b$_m$’s are the relative strengths of various partial waves contributing to the final states associated with $\Sigma$ and $\Pi$ symmetries, respectively. The $\delta_l$’s expressed in radian signify the phase differences concerning the lowest order partial wave responsible for the transition. Lastly, each m used in this caption is nothing but the whole numbers.}
\label{table:low_energy_band}
\vspace{2mm}
\begin{tabular}{cccccc}

\hline
\hline
   \textbf{\vtop{\hbox{\strut Symmetries of the}\hbox{\strut transitional states }}\hspace{5mm}}    & \textbf{\vtop{\hbox{\strut \ \ \  Weighting ratio of}\hbox{\strut \ \ \ \ the partial waves}\hbox{\strut $\hspace{6mm}a_0 : a_1 : a_2$ for $\Sigma$}\hbox{\strut $\hspace{6.5mm}b_1 : b_2  : b_3$ for $\Pi$}}\hspace{5mm}}    & \textbf{\vtop{\hbox{\strut   Phase differences}\hbox{\strut \ \ \ \ \ \ (radian) }\hbox{\strut \   $\delta_{s-p}$, $\delta_{s-d}$ for $\Sigma$}\hbox{\strut  \  $\delta_{p-d}$, $\delta_{p-f}$ for $\Pi$}}\hspace{5mm}}  & \textbf{\vtop{\hbox{\strut Adjusted }\hbox{\strut $R^{2}$ value }\hbox{\strut \ \ \ \ \ \ \ \ \ }\hbox{\strut \ \ \ \ \ \ \ \ \ }}\hspace{5mm}}   & \textbf{\vtop{\hbox{\strut$R^{2}$ value}\hbox{\strut }}\hspace{5mm}}\\
\hline

$\Sigma$ $\longrightarrow$ $\Sigma$ &2.26 : 0.68 : 3.08    & 1.16, -0.26  & 0.948 & 0.952\\
\hline
$\Sigma$ $\longrightarrow$ $\Pi$ &1.25 : 1.99 : 0.69    & 1.43, 0.99  & 0.752 & 0.772\\
\hline

\multirow{2}{*}{$\Sigma$ $\longrightarrow$ $\Sigma$ $+$ $\Pi$ } &1.31: 0.50 : 0.44   & 0.41, 0.63    & \multirow{2}{*}{0.992} & \multirow{2}{*}{0.994}\\
                       &0.92 : 1.61 : 1.78         & 0.13, 2.28      &  & \\
\hline
\hline
\end{tabular}
\end{table*}

Let's look at the kinetic energy distributions as shown in Fig. \ref{fig:co_ke_dea}. It is obvious to notice that anions with lower momentum contribute more to the intensity distribution in the case of the parallel slicing method due to the inclusion of the entire Newton spheres having diameter $\le$ 50 ns. The wedge slicing method strongly diminishes this exaggerated nature in the lower momentum range for parallel slicing due to a nearly equal fraction contribution with unequal momenta to the kinetic energy distribution. At 10.5, 11.0 and 11.5 eV incident energies in Fig. \ref{fig:co_ke_dea} (b), the two most probable peak positions for O$^-$ ions are distinguishable but are not well resolved for 10.5 and 11 eV due to the insufficient resolution of the electron beam. It also reveals that the intensity of the lower energy peak is comparably small concerning the higher energy peak. Thus, while analyzing the intensity of nascent fragments, comparing two distinct kinetic energy peaks for particular beam energy is scientifically meaningless if one uses a kinetic energy distribution extracted from time-gated parallel slices. Kinetic energy distribution of O$^-$/CO for 11.5 eV incident electron energy is fitted using Gauss model as shown in Fig. \ref{fig:ke_fit}.  The fit reveals fragmented anions having kinetic energy near 0.1 and 0.5 eV is most probable, which subsequently demands the existence of two dissociation channels as reported by Nag \textit{et al.} \cite{nag2015fragmentation}. Since energy-momentum conservation for the DEA dynamics signifies that lower kinetic energy of the fragmented anions directly maps to the dissociation channel with higher threshold energy, anions with the most probable kinetic energy at 0.1 eV should correspond to the highly energetic dissociation channel. The fit as shown in Fig. \ref{fig:ke_fit} also reflects the dissociation channel having less threshold energy is more probable to generate O$^-$ fragments. However, the ion intensity is limited (close to 0) at the exact center in each solid angle-shaped wedge slice image.  We have drawn the histogram plot by binning the x-y plane. The zero-energy bin must contain some ion counts rather than exact zero intensity as the kinetic energy of the fragments can not be zero if the incident electron has some energy. Let us consider that an incident electron having energy E is resonantly captured by a diatomic molecule AB to form a temporary negative ion (TNI) (AB$^-)^*$, and subsequently dissociates into a neutral fragment A and a negative ion fragment B$^-$. The total energy available to dissociation is given by, $\Delta{E} = (E-E_{T_h})$, where, $E_{T_h}$ is the threshold energy required for dissociation. If the velocity of the A and B$^-$ nascent fragments are $v_A$ and $v_B$, respectively, then using the energy-momentum conservation principle,
the kinetic energy of B$^-$ fragments:
\begin{equation} \label{eq1}
\begin{split}
    K_B & = (\frac{m_A}{m_A+m_B})\Delta{E}\\
\end{split}
\end{equation}

\begin{table*}
 \centering

\caption{The model parameters are reported here for the angular distribution of O$^-$ ions with kinetic energy 0.3 to 1.2 eV at 10.5 eV incident electron energy. The a$_m$’s and b$_m$’s are the relative strengths of various partial waves contributing to the final states associated with $\Sigma$ and $\Pi$ symmetries, respectively. The $\delta_l$’s expressed in radian signify the phase differences concerning the lowest order partial wave responsible for the transition.}
\label{table:high_energy_band}
\vspace{2mm}
\begin{tabular}{cccccc}

\hline
\hline
   \textbf{\vtop{\hbox{\strut Symmetries of the}\hbox{\strut transitional states }}\hspace{5mm}}    & \textbf{\vtop{\hbox{\strut \ \ \  Weighting ratio of}\hbox{\strut \ \ \ \ the partial waves}\hbox{\strut $\hspace{6mm}a_0 : a_1 : a_2$ for $\Sigma$}\hbox{\strut $\hspace{6.5mm}b_1 : b_2  : b_3$ for $\Pi$}}\hspace{5mm}}    & \textbf{\vtop{\hbox{\strut   Phase differences}\hbox{\strut \ \ \ \ \ \ (radian) }\hbox{\strut \   $\delta_{s-p}$, $\delta_{s-d}$ for $\Sigma$}\hbox{\strut  \  $\delta_{p-d}$, $\delta_{p-f}$ for $\Pi$}}\hspace{5mm}}  & \textbf{\vtop{\hbox{\strut Adjusted }\hbox{\strut $R^{2}$ value }\hbox{\strut \ \ \ \ \ \ \ \ \ }\hbox{\strut \ \ \ \ \ \ \ \ \ }}\hspace{5mm}}   & \textbf{\vtop{\hbox{\strut$R^{2}$ value}\hbox{\strut }}\hspace{5mm}}\\
\hline

$\Sigma$ $\longrightarrow$ $\Sigma$ &1.97 : 0.69 : 0.27    & 1.57, 2$\times$ 10$^{-5}$  & 0.847 & 0.861\\
\hline
$\Sigma$ $\longrightarrow$ $\Pi$ &1.24 : 1.74 : 0.99    & 1.19, 1.29 & 0.935 & 0.941\\
\hline

\multirow{2}{*}{$\Sigma$ $\longrightarrow$ $\Sigma$ $+$ $\Pi$ } &1.17: 0.36 : 0.25   & 1.14, 0.84    & \multirow{2}{*}{0.992} & \multirow{2}{*}{0.993}\\
                       &1.22 : 1.20 : 1.76         & 0.63, 2.08      &  & \\
\hline
\hline
\end{tabular}
\end{table*}

\begin{figure*}
\centering
\includegraphics[scale=0.9]{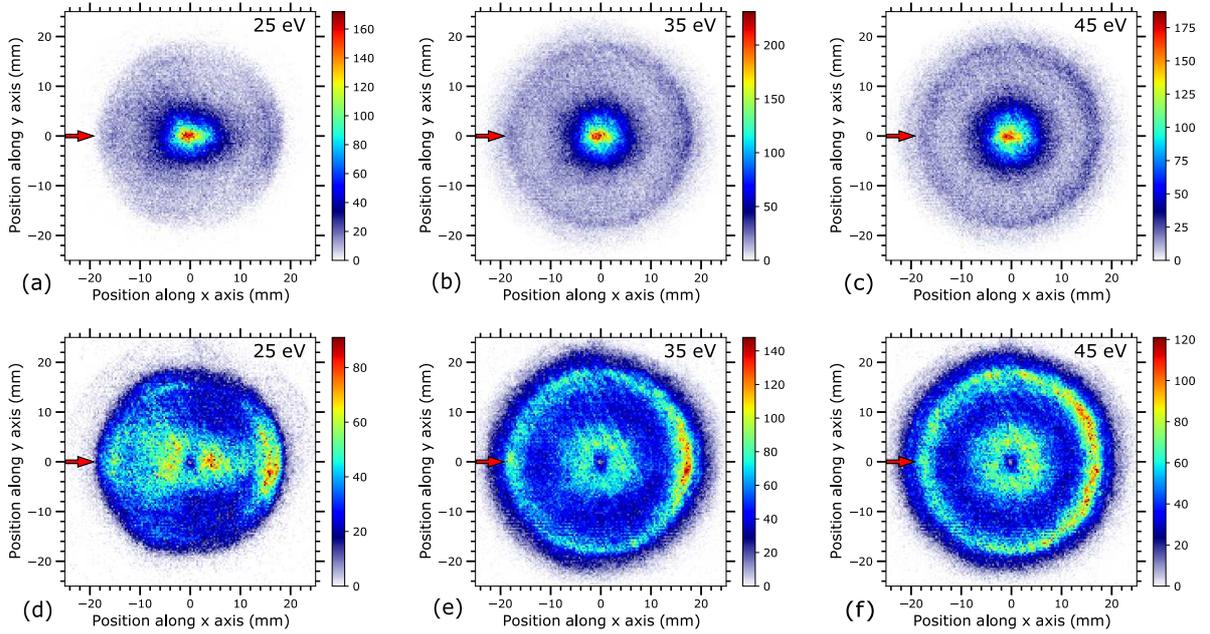}
\caption{\footnotesize{a,b and c are the 50 ns time slice images of O$^{-}/$CO at 25, 35 and 45 eV incident electron energies, respectively. d,e and f are the wedge slice images of O$^{-}/$CO at 25, 35 and 45 eV incident electron energies, respectively. The red arrow indicates the electron beam direction.}}
  \label{fig:covsi_polar}
\end{figure*}
\begin{figure*}%
    \centering
    \includegraphics[scale=0.85]{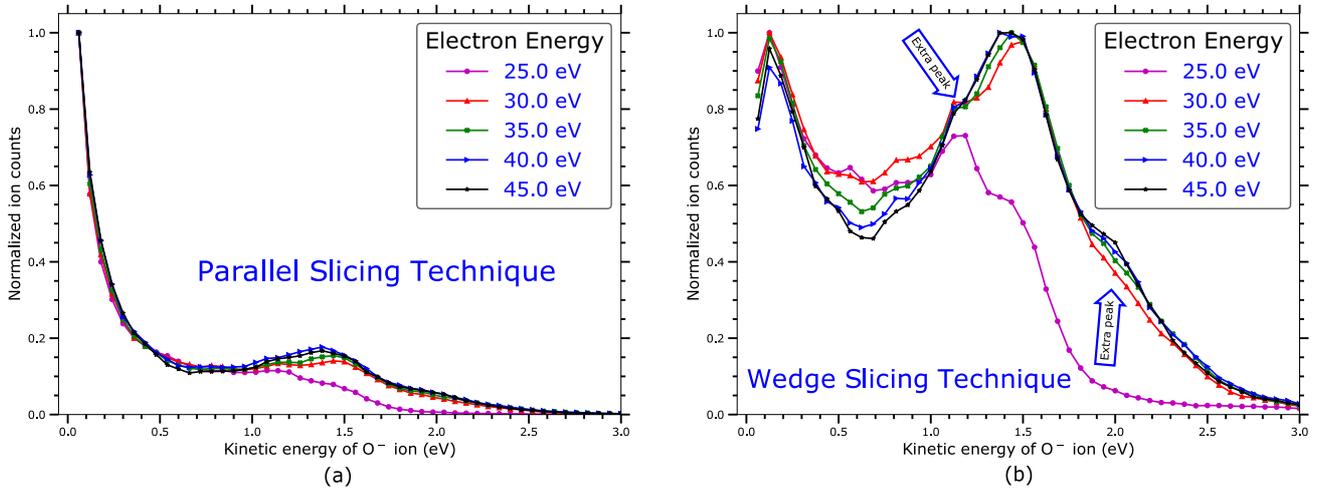}
  \caption{\footnotesize{Kinetic energy distributions of O$^-$/CO are plotted as a function of incident electron energy for (a) parallel slicing technique and (b) wedge slicing technique. The annotation-`extra peak' indicates peaks not clearly visible in the parallel slicing method.}}
  \label{fig:ke_all_polar}
\end{figure*}
From eq. \ref{eq1}, we can see that the kinetic energy of the fragments can not be zero unless the incident electron has energy (E$_{Th}$). It is true not only for DEA but also for all fragmentation dynamics. Let's consider the kinetic energy distribution in DEA dynamics for O$^{-}/$CO as shown in Fig. \ref{fig:co_ke_dea}. We observe a shift in most probable peak positions within reasonable limits, i.e., how much the peak position departs in a parallelly time-gated kinetic energy distribution from its wedge one. Here, we have observed that the corresponding peak position difference between the two different techniques is about 50 to 100 meV. Thus, probing electronic state properties will not diverge so much using the parallel slicing distributions.
\begin{figure*}
\centering
   {\includegraphics[scale=0.95]{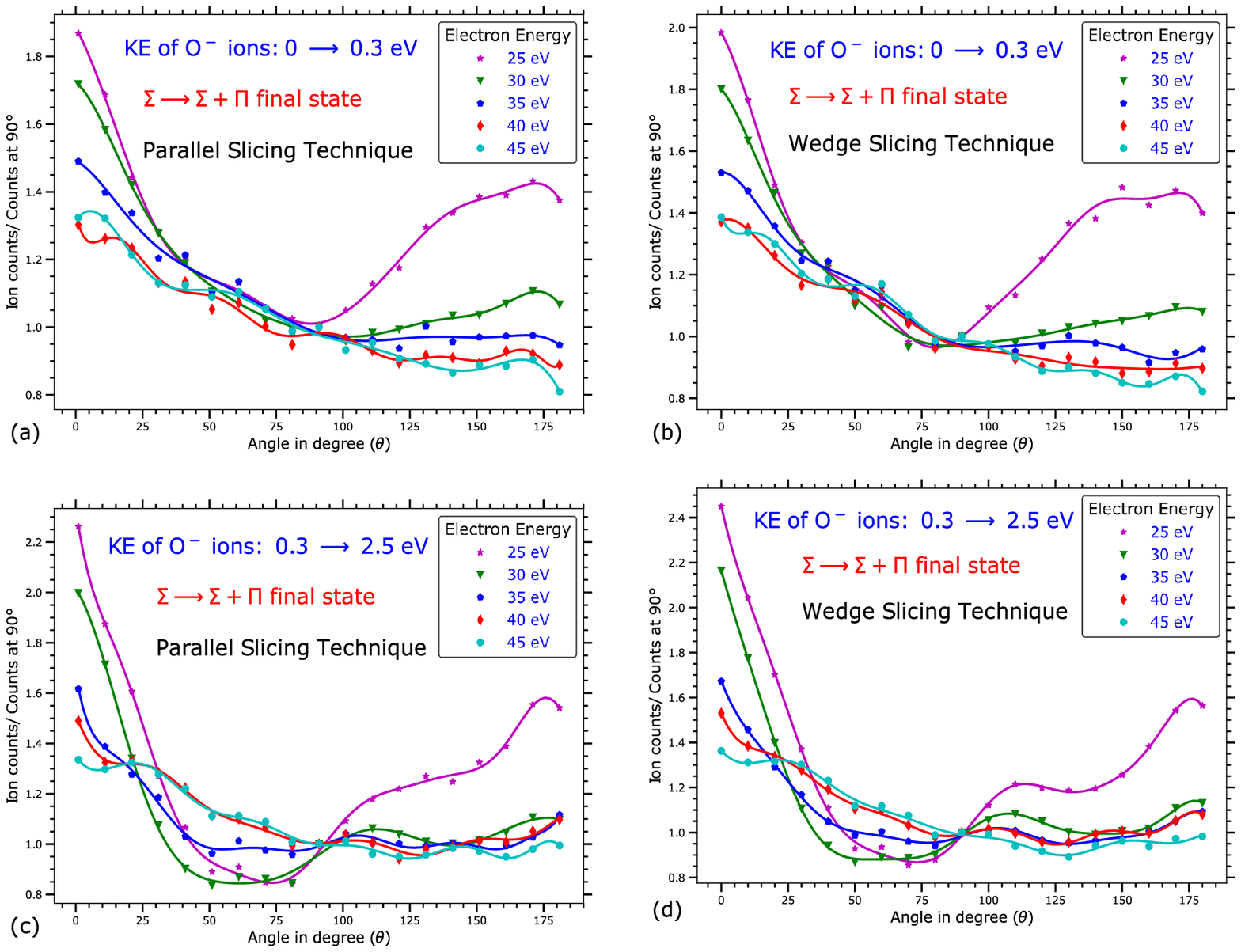}}
\caption{\footnotesize{(a,b) Angular distributions of O$^-$ ions with kinetic energy of 0.0 to 0.3 eV are plotted as a function of incident electron energy for parallel slicing technique and wedge slicing technique. (c,d) Angular distributions of O$^-$ ions with kinetic energy of 0.3 to 2.5 eV are plotted as a function of incident electron energy for parallel slicing technique and wedge slicing technique. Time slice width is 50 ns for each data. All the angular distributions have been fitted using the Van Brunt \cite{van1974breakdown} model for initial state associated with $\Sigma$ symmetry to final state associated with the addition of $\Sigma$ and $\Pi$ symmetry.}}
  \label{fig:co_ad_polar_low}
\end{figure*}
\begin{figure*}
    \centering
    \includegraphics[scale=0.85]{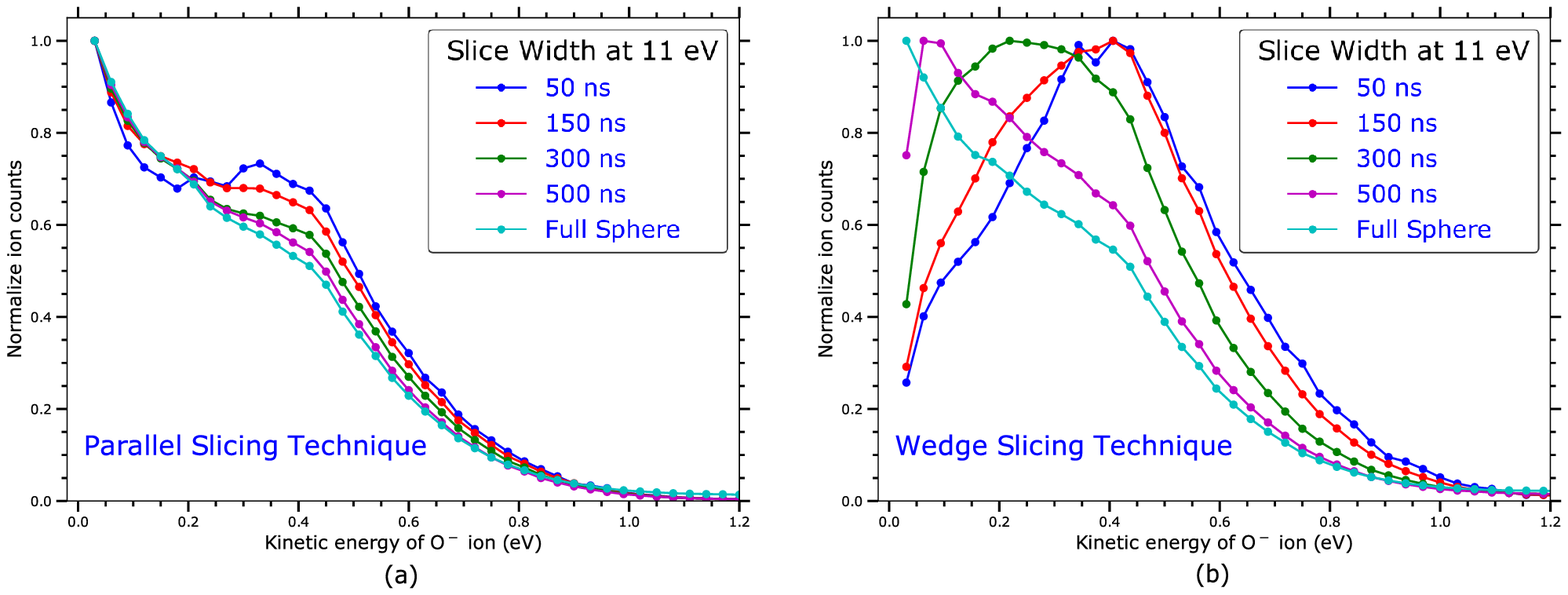}
  \caption{\footnotesize{kinetic energy distributions of O$^-$/CO at 11 eV incident electron energy are plotted as a function of time slice width for (a) parallel slicing technique  and (b) Wedge slicing technique.}}
  \label{fig:slice_va_ke_co}
\end{figure*}

We have extracted the angular distribution of O$^-$/CO having kinetic energy 0.0 to 1.2 eV using wedge slicing as well as parallel slicing method as shown in Fig. \ref{fig:co_ad_flat_wedge_all}. The lower momentum ions' exaggerated nature in parallel slicing is not manifested in the angular distributions extracted using the wedge slicing technique. The reasons behind similar results can be explained using the dependency of angular distribution function on the radius of the Newton sphere (r), azimuth ($\theta$ ) and polar angle ($\phi$). The anisotropy magnitude in angular distribution for DEA dynamics given by O'Malley and Taylor \cite{o1968angular} directly depends on azimuth angle ($\theta$) and is independent of the radius of the Newton sphere (r) and polar angle ($\phi$). This feature grounds from the definition of angular momentum where the radial component of linear momentum does not contribute to it. In addition, for 10.5 eV incident electron energy, we have determined the angular distribution of O$^-$ ions for two different kinetic energy bands; one from 0.0 to 0.3 eV another from 0.3 to 1.2 eV. Coincidentally, both the distributions extracted using the wedge slicing method follow an almost similar anisotropy nature as shown in Fig. \ref{fig:ad_co_10_5_fit}. In the same figure, the distributions have been fitted using the model of O'Malley and Taylor \cite{o1968angular} as written in eq. \ref{eq2} and indicate that the final states of each energy band are associated with both the $\Sigma$ and $\Pi$ symmetries.

\begin{equation} \label{eq2}
    \rm{I}(k, \theta, \phi) \approx \left\lvert \sum_{L = \mid \mu \mid}^{\infty} a_{L, \mid \mu \mid}(k)  Y_{L, \mu}(\theta, \phi)\right\rvert^{2}
\end{equation}

The significance of each symbol used in this eq. \ref{eq2} is discussed briefly in our previous report \cite{nag2015fragmentation}. The fit parameters (similar to previous report \cite{nag2015fragmentation}) as shown in the table \ref{table:low_energy_band} and table \ref{table:high_energy_band} determined from the wedge slicing method signify O$^-$ions within lower kinetic energy band (0.0 to 0.3 eV) are dominantly associated to final state with $\Sigma$ symmetry  over $\Pi$ whereas  O$^-$ ions within higher kinetic energy band (0.3 to 1.2 eV) are dominantly associated to final states with $\Pi$ symmetry over $\Sigma$. From the fit in Fig. \ref{fig:ad_co_10_5_fit}, it is noticeable that the weakly contributing forward lobe to the angular distribution mainly originates from the final state associated with  $\Pi$ symmetry.

\begin{figure*}
    \centering
    \includegraphics[scale=0.85]{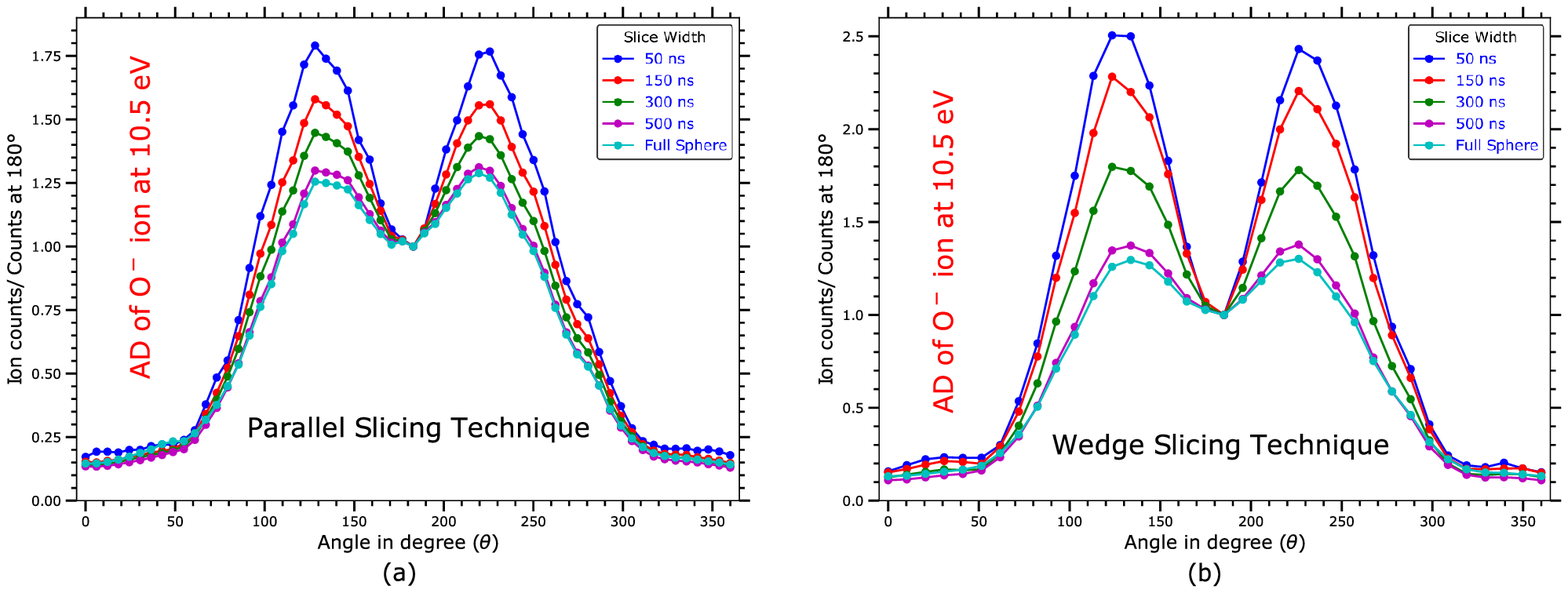}
  \caption{\footnotesize{ Angular distributions of O$^-$/CO at 10.5 eV incident electron energy are plotted as a function of time slice width for (a) parallel slicing technique  and (b) Wedge slicing technique.}}
  \label{fig:slice_variation_ad_co}
\end{figure*}

\begin{figure}
  \includegraphics[scale=0.5]{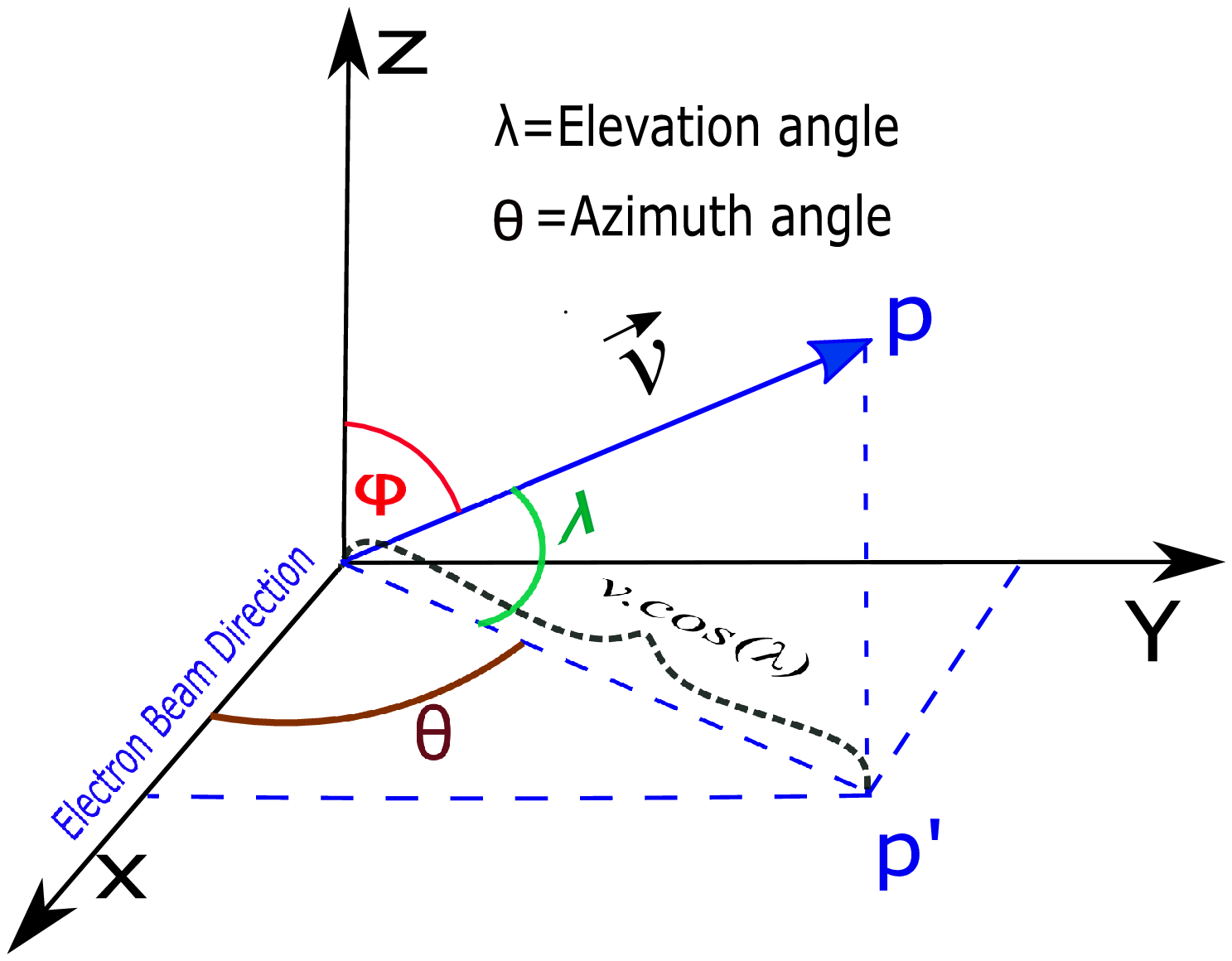}
  \caption{\footnotesize{Schematic representation for analysing the converted Newton sphere.}}
  \label{fig:explanation_slice_variation}
\end{figure}

\subsection{Dipolar dissociation to carbon monoxide}
The wedge slicing technique is further applied for analyzing the dipolar or ion-pair dissociation (DD/IPD) to carbon monoxide (CO). Fig. \ref{fig:covsi_polar} shows the momentum images in the x-y plane using both the slicing techniques for the IPD process of carbon monoxide at 25, 35 and 45 eV incident electron energies. The wedge slice images (Fig. \ref{fig:covsi_polar}) of IPD dynamics in contrast to DEA, exhibit a constant anion intensity throughout the Newton sphere. This feature may reveal the existence of a predissociation phenomenon in IPD dynamics of O$^{-}/$CO. The two intense kinetic energy bands in Fig. \ref{fig:ke_all_polar} superimpose the predissociation data. During the study of electron bream induced IPD dynamics of O$_2$, Nag \textit{et al.} \cite{nag2018study} reported the same. Such a report may be inconclusive, visualizing the momentum images extracted using time-gated parallel slices techniques. In the case of DEA resonances, we have never noticed a Newton sphere containing anions fragments distributed continuously. Thus, using such an experimental observation, one can differentiate between the DEA resonances and non-resonance IPD phenomenon originating from the ionization continuum. By studying the IPD in H$_2$, HD \& D$_2$, Chupka \textit{et al.} \cite{chupka1975high} reported the superposition of structured continuum with characteristic window resonances; which belongs to a specific Rydberg series that later experiences predissociation as well as autoionization. The wedge slicing technique to the IPD dynamics of O$^{-}/$CO strongly supports the superposition of structured continuum with two characteristic window resonances, one from 0.0 to 0.5 eV and another from 0.5 to 3.0 eV. The kinetic energy distributions of O$^-$/CO in the IPD range are extracted using both parallel as well as wedge slicing techniques as shown in Fig. \ref{fig:ke_all_polar}. The distribution reflects the two most probable peak positions for O$^-$ fragments with kinetic energy near 0.1 eV and 1.5 eV. The distribution in Fig. \ref{fig:ke_all_polar} (b) also reveals two extra peak positions for O$^-$ fragments with kinetic energy near 0.8 and 1.9 eV, as shown using the annotations with arrows. These extra peaks may indirectly map 3D oscillation in potential energy surfaces. The existence of more ion-pair dissociation channels may lead to such a phenomenon. Detailed theoretical calculations on IPD dynamics to CO can demonstrate these findings, which is extremely computationally challenging for the scientific community. Thus, wedge slicing offers more information on the kinetic energy distribution over the conventional time-gated parallel slicing method, irrespective of dynamics studies through DEA or IPD. Fig. \ref{fig:co_ad_polar_low} represent the angular distributions of O$^-$ ion for two kinetic energy bands (0.0 to 0.3 eV and 0.3 to 2.5 eV) in IPD range extracted using both the mentioned techniques. All the angular distributions are fitted using the expression given by Van Brunt \cite{van1974breakdown} and are already discussed in detail in our previous report \cite{chakraborty2016dipolar}. The best fit reveals that the final states of O$^-$ fragments are associated with $\Sigma$ and $\Pi$ symmetries as shown in Fig. \ref{fig:co_ad_polar_low}.\\

\subsection{Variation of the time slice width}
Additionally, we have studied the effect of time slice width variations for parallel slicing technique and solid angle-shaped wedge slicing. The kinetic energy distribution of O$^-$/CO at 11 eV incident beam energy and angular distributions of O$^{-}$/CO at 10.5 eV incident beam energy are shown in Fig. \ref{fig:slice_va_ke_co} and Fig. \ref{fig:slice_variation_ad_co} for different time slice width. Here the slice width in a wedge slice implies the maximum TOF interval considered along the TOF or Z-axis. The intensity of the higher energy peaks decreases with the increment of the time slice width irrespective of the method used. We have further analyzed the angular distribution of fragmented anions using both the techniques as a function of time slice width as shown in Fig. \ref{fig:slice_variation_ad_co}. Both the distributions reveal that the anisotropy of angular distribution gradually decreases with the increment of time slice width. This astonishing result in angular distribution \& kinetic energy distributions appears due to the two factors. The first factor is the VMI condition which is not maintained throughout the detected Newton sphere. Only the nascent fragments with velocity vector parallel to the detector plane, i.e., the fragment anions near the z$=0$ plane, properly maintain the VMI condition. A major and second one is that the projections of these nascent fragments to the detector are considered in the techniques (not an exact 3D bin consideration to construct the entire Newton sphere). Using the schematically drawn Fig. \ref{fig:explanation_slice_variation}, we can explain the reasons behind the appearance of this major factor. Here, the electron beam is considered along the x-axis, $\vec{v}$ is the velocity vector of the nascent fragment anions, $\lambda$ is the elevation angle concerning the z=0 plane and  $\theta$ is the angle between the electron beam and the projection of the velocity vector on the x-y plane, known as azimuth angle. Thus, we can see that $\lambda$ $\neq$ $\theta$ and our angular distributions are extracted concerning the azimuth angle $\theta$, i.e., the projection of a particular velocity vector on the x-y plane is contributing to the kinetic energy or angular distribution. Conventionally, angular distribution should be determined concerning the angle formed between the velocity vector and beam direction.
Thus, when we increase the wedge slice width, elevation angle $\lambda$ will increase, and $\phi$ will decrease accordingly. So, in this bounded geometry segment ($-\pi/2 < \lambda < \pi/2$), the weightage of the $\cos{\lambda}$ factor to the angular distribution should be a rationale for observing a decrease in anisotropy with the increment of wedge slice width. Another minor factor is the time-to z-axis scaling parameter, used to form the Newton sphere. Moradmand \textit{et al.} \cite{slaughter_2013} point out that this scaling factor is non-linear, but we have taken it as a linear one. For kinetic energy distributions extracted using the parallel slicing method as shown in Fig. \ref{fig:slice_va_ke_co}, the intensity of the high energetic peak gradually decreases with the increment of time-gated parallel slice width. Still, the peak position for the highly energetic fragments is blue-shifted slowly due to its exaggerated nature of ions with a lower momentum. At the same time, the peak shift is less in the wedge method. Both the kinetic energy distributions for an entire sphere are the same demand this data analysis by projecting it on a plane. 

\section{Conclusions} 
We have successfully implemented a solid angle-shaped wedge slicing technique and comparatively discussed this technique with a time-gated parallel slicing method. In our case, we found that the wedge slice with a limited time gate (up to 100 ns) is the finest way to determine the nascent fragment ions' kinetic energy and angular distribution. One can argue that the most probable peak positions of the nascent fragments were scientifically meaningless for the previously used parallel time-gated technique. Indeed, this intensity comparison will open a new canton in atomic, molecular and optical physics revealing the phenomenological effect of interferences arising in these states. The advantages of this wedge slice are that if the elevation angle is minimal, we can determine the kinetic energy distribution of the ions even though the VMI condition is not satisfied for the entire Newton sphere. Lastly, it is challenging for the community to determine the exact nature of the non-linearity to evaluate the time to the z-momentum scale factor. The predissociation process confirmed in IPD dynamics will open fascinating properties to study heavy, and ultralong Rydberg system \cite{hrs_review,kirrander2018heavy,minima_ulrs}. We illustrate the employment of a three-dimensional (x,y,t) MCP-Delay Line coupled detection system that can be implemented in all sorts of ion imaging in which inversion techniques are unfeasible, i.e., the breakdown of axial symmetry \cite{dinu2002application}. Thus, phenomenally correct estimates, i.e., the degree of accuracy, depending on the techniques one utilizes.

\section*{Acknowledgements} 
N.K. gratefully acknowledges the financial support from `DST of India' for the "INSPIRE Fellowship" program. A.P. expresses deep appreciation to the "Council of Scientific and Industrial Research (CSIR)" for the financial assistance. We gratefully acknowledge financial support from the Science and Engineering Research Board (SERB) for supporting this research under  Project No. “CRG/2019/000872”. N.K. thanks, Pamir Nag, for his kind suggestions and fruitful discussions.\\

\bibliography{wedge_slicing} 

\end{document}